\documentclass[twocolumn,showpacs,preprintnumbers,superscriptaddress,amsmath,amssymb]{revtex4}

\usepackage[dvips]{color}

\usepackage{graphicx}
\usepackage{dcolumn}
\usepackage{bm}
\usepackage{latexsym}

\begin{document}

%

\title{Enhancement of the superconducting critical temperature in Bi$_2$Sr$_2$CaCu$_2$O$_{8+\delta }$ by controlling disorder outside CuO$_2$ planes}

\author{H. Hobou}
 \affiliation{Department of Physics, University of Tokyo, Tokyo 113-0033, Japan}
\author{S. Ishida}
 \affiliation{Department of Physics, University of Tokyo, Tokyo 113-0033, Japan}
\author{K. Fujita }
\affiliation{Department of Advanced Materials Science, University of Tokyo, Tokyo 113-0033, Japan}

\author{M. Ishikado}
\affiliation{Department of Physics, University of Tokyo, Tokyo 113-0033, Japan}

\author{K. M. Kojima}
\affiliation{Department of Physics, University of Tokyo, Tokyo 113-0033, Japan}
\author{H. Eisaki}
\affiliation{Nanoelectronics Research Institute, National Institute of Advanced Industrial Science and Technology (AIST), Tsukuba 305-8568, Japan}
\author{S. Uchida}
\affiliation{Department of Physics, University of Tokyo, Tokyo 113-0033, Japan}
\affiliation{Department of Advanced Materials Science, University of Tokyo, Tokyo 113-0033, Japan}


{\color{red}}
\begin{abstract}

We investigate the effect of disorder at various lattice sites outside the CuO$_2$ plane on the superconducting critical temperature $T_c$ of the cuprate superconductor Bi$_2$Sr$_2$CaCu$_2$O$_{8+\delta }$ (Bi2212).
The most effective disorder turns out to be that at the Sr site in the neighboring blocks, which contain apical-oxygen atoms. The Sr-site disorder reduces $T_c$ and also produces a residual component ($\rho _0$) in the in-plane resistivity, the magnitude of which is found to be proportional to the magnitude of $T_c$ reduction. We demonstrate that both $T_c$-degradation rate and $\rho _0$ decrease as the number of the CuO$_2$ planes increases in the unit cell, that is, as the number of the neighboring SrO blocks decreases. In real crystals of Bi-based cuprates,
main source of disorder is Bi atoms randomly occupying the Sr sites. We have succeeded in reducing the Bi content at Sr-sites as much as possible and achieved $T_c$=98K, a high-$T_c$ record in Bi2212.

\end{abstract}


\maketitle

\section{\label{sec:level1}Introduction}

Sixteen years have passed, since the maximum superconducting critical temperature $T_c \sim $135K was recorded in 1993~\cite{schilling}. Although many researchers have tried to raise $T_c$, no successful results have yet been reported and the question of which factors determine $T_c$ in cuprates has not been fully understood. In spite of having the same CuO$_2$ structural unit and similar structure, $T_c$ of cuprates is strongly material dependent, ranging from $\sim$ 30 to 135K, indicating that $T_c$ is not determined by a single parameter or a few parameters which characterize the electronic or phononic systems. So many parameters have so far been known to affect $T_c$, e.g., (i) doping level, (i\hspace{-.1em}i) superfluid density~\cite{uemura}, (i\hspace{-.1em}i\hspace{-.1em}i) number of CuO$_2$ planes in a unit cell~\cite{iyo2}, (i\hspace{-.1em}v) disorder in the CuO$_2$ plane~\cite{fukuzumi}, and (v) distance of the apical-oxygen atoms from the CuO$_2$ plane~\cite{pavarini}. Eisaki \textit{et al.}~\cite{eisaki} pointed out that disorder caused by dopant atoms outside the CuO$_2$ planes has also influence on $T_c$. As a good demonstration, $T_c$ is enhanced to 95K for a single-layer cuprate Sr$_2$CuO$_{3+\delta }$ when randomly distributed dopant atoms, apical-oxygen atoms in this material, form an ordered array~\cite{liu}. Fujita \textit{et al.}~\cite{fujita} quantitatively examined the effect of this type of disorder in the single-layer system Bi$_{2.0}$Sr$_{1.6}$Ln$_{0.4}$CuO$_{6+\delta }$ (Ln-Bi2201) in which disorder is controllably introduced by the partial substitution of the rare earth element (Ln) for Sr and the degree of disorder is varied by changing Ln with different ionic radii~\cite{ahrens}. They showed that $T_c$ appreciably decreases with increasing degree of disorder at Sr sites.

\indent In this paper we focus on the bi-layer Bi$_2$Sr$_2$CaCu$_2$O$_{8+\delta }$ (Bi2212), and examine the effect of disorder at various sites outside the CuO$_2$ planes (out-of-plane disorder). Based on this result we explore methods of enhancing and maximizing $T_c$ of Bi2212. In addition, by comparison with the available data on single-layer and tri-layer cuprates, we elucidate how the reduction in $T_c$ due to out-of-plane disorder depends on the number of CuO$_2$ planes in unit cell.

\indent The crystal structure of Bi2212 is composed of three building blocks other than two CuO$_2$ planes: (1) Bi$_2$O$_2$ block, (2) SrO block, and (3) Ca block. These three blocks are charge reservoirs and, at the same time, sources of disorder. The excess oxygen atoms are located in or near the Bi$_2$O$_2$ block which act as hole dopant. A fairy large amount of Pb can replace Bi which acts to suppress the structural supermodulation of the BiO layers. Bi$^{3+}$ ions tend to replace the Sr$^{2+}$ ions in the SrO block (sometimes called as ``site mixing''). The Ca$^{2+}$ ions in the plane sandwiched by two CuO$_2$ planes can easily be replaced by rare-earth (Y or Ln$^{3+}$) ions with relatively small ionic radii. Bi$^{3+}$ and Ln$^{3+}$ substituted for Sr$^{2+}$ and Ca$^{2+}$ decrease the hole density in the CuO$_2$ plane. Normally grown single crystals of Bi2212 contain these three types of disorders. In this regard Bi2212 is a highly disordered system, but is a good ground for studying and discriminating the effects of disorder at different sites, on superconducting $T_c$. Usually increasing or decreasing disorder in one particular block changes the doping level in the CuO$_2$ planes, but in Bi2212 the doping level can be controlled also by the other two blocks. This gives an opportunity to examine pure disorder effect from individual block without changing the doping level upon which $T_c$ depends very sensitively. 

\begin{table*}[t!]
\caption{\label{tab:table1}ICP-analyzed compositions and maximum $T_c$ (onset) of the crystals (most of them containing Pb) in this study. The molar ratio of Cu composition is set to 2.}
\begin{ruledtabular}
\begin{tabular}{lcl}
Nominal composition&Compositions (ICP)&$T_c^{max}$\\
&Bi / Pb / Sr / Ca / Y / Cu &(K)\\
\hline
Bi$_{2.2}$Sr$_{1.8}$CaCu$_2$O$_{8+\delta}$&2.16/0.00/1.85/1.06/0.00/2 & 92.0\\
Bi$_{2.0}$Sr$_{2.0}$Ca$_{0.92}$Y$_{0.08}$Cu$_2$O$_{8+\delta}$&2.02/0.00/1.98/0.85/0.08/2 & 96.0\\
Bi$_{1.6}$Pb$_{0.4}$Sr$_2$CaCu$_2$O$_{8+\delta}$&1.69/0.33/1.96/0.92/0.00/2 & 96.0\\
Bi$_{1.4}$Pb$_{0.8}$Sr$_2$CaCu$_2$O$_{8+\delta}$&1.42/0.55/1.97/0.96/0.00/2 & 97.5\\
Bi$_{1.3}$Pb$_{0.9}$Sr$_2$CaCu$_2$O$_{8+\delta}$&1.31/0.66/1.94/1.08/0.00/2 & 94.0\\
Bi$_{1.6}$Pb$_{0.4}$Sr$_2$Ca$_{0.9}$Y$_{0.1}$Cu$_2$O$_{8+\delta}$&1.58/0.36/1.92/0.93/0.11/2 & 94.0\\
Bi$_{1.6}$Pb$_{0.4}$Sr$_2$Ca$_{0.95}$Y$_{0.05}$Cu$_2$O$_{8+\delta}$&1.50/0.37/1.97/0.92/0.05/2 & 96.0\\
Bi$_{1.4}$Pb$_{0.7}$Sr$_2$Ca$_{0.97}$Y$_{0.03}$Cu$_2$O$_{8+\delta}$&1.33/0.65/2.00/0.98/0.03/2 & 98.0\\
\end{tabular}
\end{ruledtabular}
\end{table*}

\indent We show below that the disorder in the SrO block is most influential on $T_c$, while disorder in other two blocks has very small effect on $T_c$. In real crystals of Bi2212 a certain fraction of Bi ions occupy the Sr sites in the SrO block which are dominant source of disorder that reduces $T_c$. Therefore, the reduction in Bi content in the SrO block is an effective way to increase $T_c$. In the present work we have succeeded in reducing Bi content as much as possible and increasing $T_c$ up to 98K, a record value in Bi2212. Finally, we illustrate that the $T_c$ degradation due to this type of disorder weakens as the number of CuO$_2$ planes in a unit cell increases, which explains why single-layer system is most strongly material dependent.


\section{\label{sec:level2}Experimental methods}

High quality single crystals of Bi2212 were grown using the traveling-solvent-floating-zone (TSFZ) method. In this study, we made the samples (1) Bi$_{2+x}$Sr$_{2-x}$CaCu$_2$O$_{8+\delta }$, (2) Bi$_2$Sr$_2$Ca$_{1-y}$Y$_y$Cu$_2$O$_{8+\delta }$, (3) Bi$_{2-w+x}$Pb$_w$Sr$_{2-x}$CaCu$_2$O$_{8+\delta }$, and (4) Bi$_{2-w}$Pb$_w$Sr$_2$Ca$_{1-y}$Y$_y$Cu$_2$O$_{8+\delta }$ from well-dried powders of Bi$_2$O$_3$, PbO, SrCO$_3$, CaCO$_3$, Y$_2$O$_3$, and CuO. The crystal growth was carried out in air and at growth speeds of 0.15-0.2 mm/h for Pb-free Bi2212 and 0.5mm/h for Pb-doped Bi2212. For Pb-doped crystals the actual Pb content tends to be smaller than the nominal one since Pb is easily evaporated out during the TSFZ growth. Inductively coupled plasma (ICP) spectroscopy was used for the composition analysis and a superconducting-quantum-interference-device (SQUID) magnetometer was used for measurement of $T_c$. Table \ref{tab:table1} lists both nominal and actual compositions of the crystals containing Pb in this study. $T_c$ was defined as the onset temperature at which the zero-field-cooled susceptibility starts to drop. In-plane resistivity was measured by the six-probe method with current kept constant (1mA). We measured the resistivity for more than ten samples for each composition to reduce uncertainty in the magnitude and to check homogeneity of crystals. 

$T_c$ of each sample  or the doped hole density can be varied by annealing it under different atmosphere. Oxidation annealing was performed in air or under oxygen gas flow, and deoxidation annealing was done in vacuum or under nitrogen gas flow. $T_c^max$ in Table \ref{tab:table1} is the maximum $T_c$ which was attained by changing the excess oxygen content $\delta $ under various annealing conditions. In the following, $T_c$ means $T_c^max$, if not otherwise mentioned. The annealing condition giving $T_c^max$ is attached to the caption of the result. We can vary two parameters, operating temperature and oxygen partial pressure, for annealing. From thermogravimetric analysis, we found that for Pb- and Y-doped crystals $\delta $ can be varied over a wider range than the reported one~\cite{fujii}, 0.20 $< \delta <$ 0.26 for Bi2212 without Pb or Y. In fact, we found that $T_c$ shows a maximum value when the annealing condition or atmosphere is in extreme reduction or extreme oxidation attainable in this study (a typical example is shown in Fig. \ref{fig:4}).


\section{\label{sec:level3}Experimental Results}

\subsection{Effect of out-of-plane disorder on the in-plane resistivity}

The Bi/Sr-site mixing, the dominant source of disorder in Bi2212, occurs easily at Sr site in Bi2212 due to following reasons: (1) Bi$^{3+}$ cations tend to be attracted to CuO$_2$ planes and replaced to $A$ site (nominally Sr site in Bi2212)  because the CuO$_2$  bilayer is negatively charged, [CuO$_2$-Ca-CuO$_2$]$^{2-}$, and (2) the apical-oxygen block allocating Sr$^{2+}$ at $A$ site is relatively unstable as compared with that allocating Ba$^{2+}$ because of smaller ionic radius of Sr$^{2+}$ and hence smaller tolerance factor.~\cite{mca}$^{,}$\cite{hakuraku} These make the Sr site vulnerable to the intrusion of Bi$^{3+}$. Normally, Bi2212 crystals grown by TSFZ method contain an appreciable amount of excess Bi, $x$=0.10-0.20 in the formula Bi$_{2+x}$Sr$_{2-x}$CaCu$_2$O$_{8+\delta}$, and $T_c$ is, at its highest, 92K even for optimal doping. From the previous works~\cite{eisaki,fujita}, it is likely that $T_c$ is reduced due to this Sr-site disorder.

\begin{figure}[t!]
\includegraphics[width=0.9\columnwidth ]{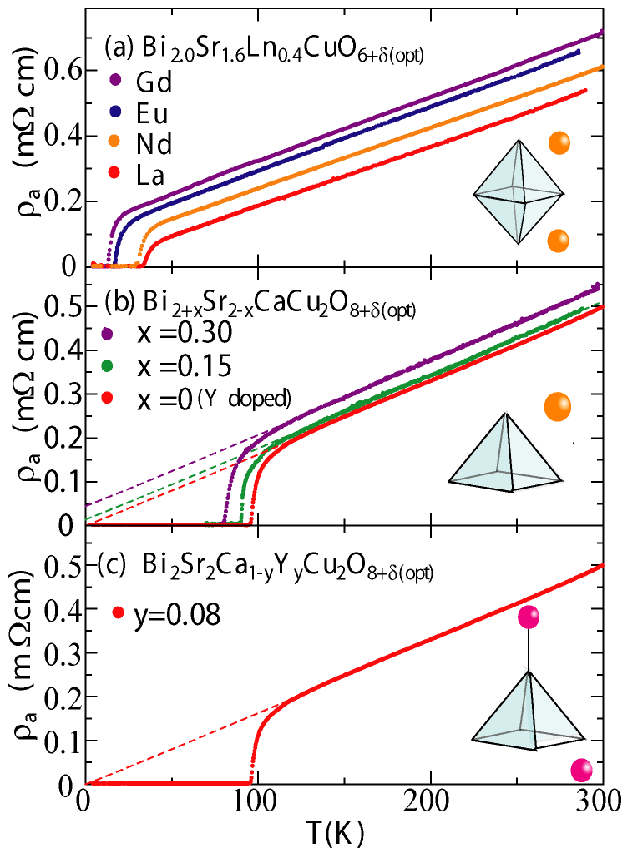}
\caption{\label{fig:1}(a)-(c) Effect of disorder at out-of-plane sites of optimally doped Bi-cuprates on the in-plane resistivity with the current flowing along $a$ axis. We define the residual resistivity component $\rho_0 $ by the extrapolation of the temperature-linear resistivity to 0K. (a) Single-layer Bi$_2$Sr$_{1.6}$Ln$_{0.4}$CuO$_{6+\delta(opt)}$ with  various rare-earth Ln substituted for Sr (Ln=Gd, Eu, Nd, and La from top to bottom). Disorder is introduced by the 20\% substituted rare-earth(Ln) element in the SrO block. The data of Ln-Bi2201 are from Ref. 8. (b) Bi$_{2+x}$Sr$_{2-x}$CaCu$_2$O$_{8+\delta(opt)}$ with disorder due to Bi atoms occupying the Sr sites ($x$=0.30, 0.15, and 0 from top to bottom). The data for $x$=0 are the same as that in (c). (c) Bi$_2$Sr$_2$Ca$_{1-y}$Y$_y$Cu$_2$O$_{8+\delta(opt)}$ (Y-Bi2212) with disorder both in the Ca and near Bi$_2$O$_2$ block. The hole density is set near optimal ($\sim$ 0.2 holes/in-plane Cu) in each sample by adjusting oxygen content. Near-optimal doping is guaranteed by the magnitude and the slope of the temperature-linear resistivity.} 
\end{figure}

\begin{figure}
\includegraphics[width=0.9\columnwidth ]{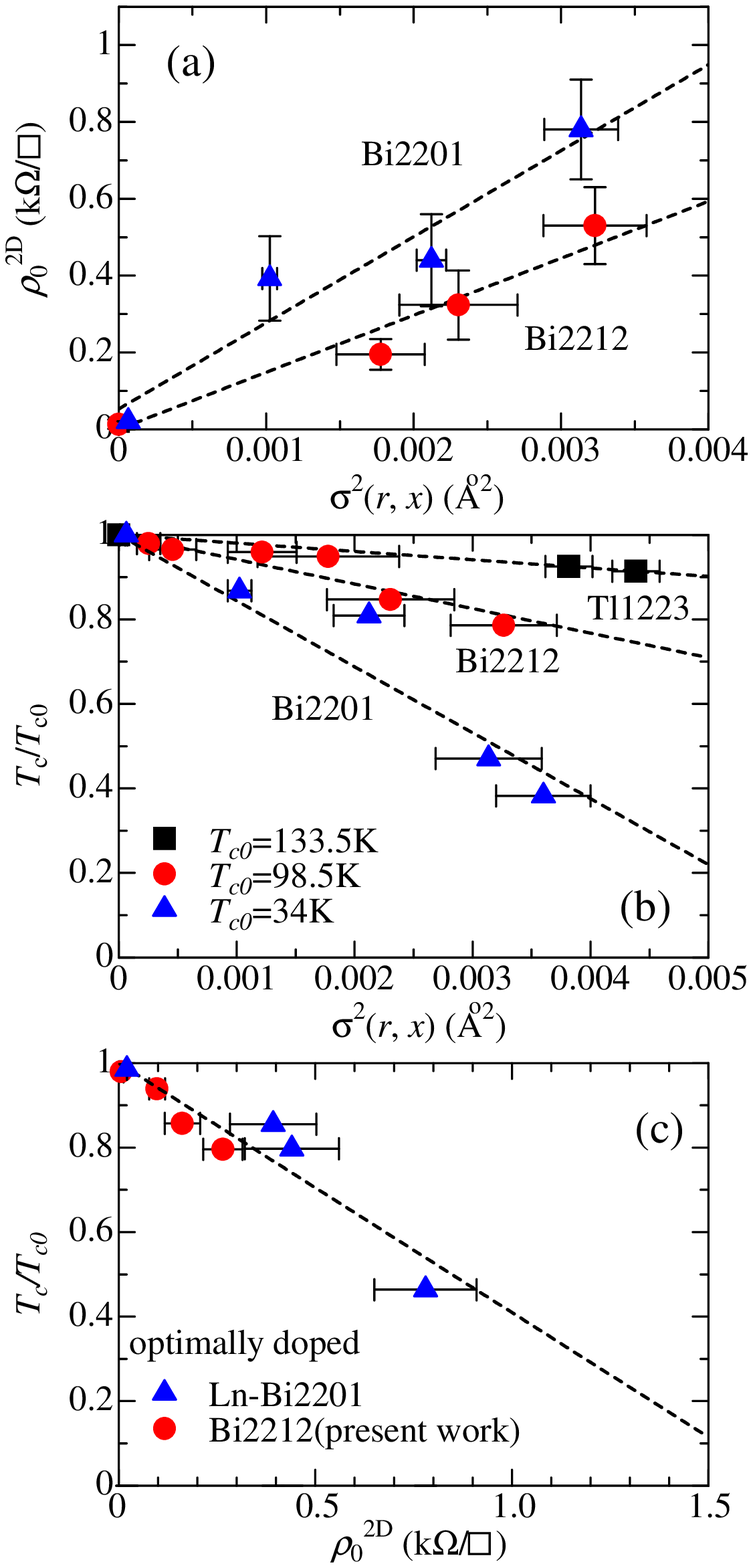}
\caption{\label{fig:2}(a) Residual resistivity per CuO$_2$ plane, $\rho _0 ^{2\rm{D}}$, vs degree of disorder defined in the text. Broken lines are the least squares fits to the data.
(b) $T_c$ normalized by $T_{c0}$ (defined in the text) plotted against $\sigma ^2 (r, x)$ for $n$=1(Bi2201) (Ref. \cite{fujita}), $n$=2 (Bi2212), and $n$=3 (Tl1223) (Ref. \cite{iyox}).
(c) $T_c$ degradation against two-dimensional residual resistivity (per CuO$_2$ plane) induced by disorder.}
\end{figure}

\indent Figures \ref{fig:1}(a) and (b) show the temperature dependences of the in-plane resistivities for optimally doped Ln-Bi2201 and Bi2212 ($\sim $ 0.2 holes/Cu) with disorder in the SrO block: (a) rare-earth element Ln substitution for Sr in Bi2201, and (b) Bi/Sr-site mixing in Bi2212. The resistivity shown in Fig. 1(c) is for a sample in which the Sr-site disorder is reduced, but Ca site sandwiched by two CuO$_2$ planes is disordered due to partial Y substitution. Note that disorder due to excess oxygen atoms and structural supermodulation always exists near or inside the Bi$_2$O$_2$ block. It is demonstrated for Ln-Bi2212 that Ln disorder induces residual resistivity component $\rho _0$ in the in-plane resistivity and the magnitude of $\rho _0$ (vertical shift of the resistivity curve) is correlated well with the degree of disorder in the SrO block~\cite{fujita}. Fig. \ref{fig:1}(b) shows that this is also the case with Bi2212 with Bi in the Sr sites. In Fig. \ref{fig:2}(c) we show a plot of the $T_c$-degradation rate ($T_c$ / $T_{c0}$) against the magnitude of two-dimensional residual resistivity per CuO$_2$ plane for disorder in the SrO block in both Bi2201 and Bi2212. We see that the reduction in $T_c$ due to disorder at Sr sites roughly scales with the magnitude of $\rho _0$. Note that the disorder in the Ca block due to Y substitution and that near the Bi$_2$O$_2$ block due to excess oxygen atoms produce almost no residual resistivity. Hence, we infer that disorder in these two blocks has negligible effect on $T_c$, even though the location of Ca site is close to the CuO$_2$ planes.

\indent Further insight into the effect of out-of-plane disorder is gained by a comparison with single-layer and tri-layer cuprates. For this comparison, we use $\sigma ^2 $ (variance of ionic radii of cations at the Sr site) as a measure for the degree of disorder~\cite{attfield}. $\sigma ^2 $ is determined by two factors: one is the fraction of different cations intruding the $A$ site, $x$, and the other is the ionic radius $r$ of the cations substituted at the $A$ site --- ionic radius mismatch between intruding cations and Sr. Specifically, in the case of Bi$_2$Sr$_{2-x}$Ln$_{x}$CuO$_{6+\delta}$, $\sigma ^2 (r, x)$ is determined by 
\begin{eqnarray*}
\sigma ^2 (r, x) &=& \frac{x}{2}r^2_{\mathrm{Ln^{3+}}}+(1-\frac{x}{2})r^2_{\mathrm{Sr^{2+}}} \\
                 & &   - [\frac{x}{2}r_{\mathrm{Ln^{3+}}} + (1-\frac{x}{2})r_{\mathrm{Sr^{2+}}}]^2,
\end{eqnarray*}
where $r_{\mathrm{Ln^{3+}}}$ and $r_{\mathrm{Sr^{2+}}}$ are the radii of rare-earth Ln$^{3+}$ and Sr$^{2+}$ ions, respectively. For Bi$_{2+x}$Sr$_{2-x}$CaCu$_2$O$_{8+\delta}$, assuming that the intrusion of Bi is a major source of Sr-site disorder, 
\begin{eqnarray*}
\sigma ^2 (r, x) &=& \frac{x}{2}r^2_{\mathrm{Bi^{3+}}}+(1-\frac{x}{2})r^2_{\mathrm{Sr^{2+}}}  \\
                 & &   - [\frac{x}{2}r_{\mathrm{Bi^{3+}}} + (1-\frac{x}{2})r_{\mathrm{Sr^{2+}}}]^2,
\end{eqnarray*}
where $r_{\mathrm{Bi^{3+}}}$ is the ionic radius of Bi$^{3+}$.

\indent One sees in Fig. \ref{fig:1} that a small but finite residual resistivity appears and increases with increasing the Sr-site disorder in both Bi2201 and Bi2212.  For quantitative comparison, plotted in Fig. \ref{fig:2}(a) is a relationship between disorder-induced two-dimensional residual resistivity per CuO$_2$ plane ($\rho _0 ^{2\rm{D}}) $ and $\sigma ^2 (r, x)$ for Bi2201 and Bi2212. We see that the magnitude of $\rho _0 ^{2\rm{D}}$ for $n$=1 is roughly twice as large as that for $n$=2 at the same $\sigma ^2$, which indicates that the Sr-site disorder works as an elastic scatterer only for the carriers in the adjacent CuO$_2$ plane. That is, the difference between $n$=1 and $n$=2 is naturally explained by the fact that the number of SrO blocks adjacent to a CuO$_2$ plane is two in Bi2201 and one in Bi2212.

\indent $T_c$ degradation ($T_c$/$T_{c0}$) with increasing disorder ($\sigma ^2$) is plotted in Fig. \ref{fig:2}(b) for $n$=1, 2, and 3. Here we use the data for Tl$_{1+x}$Ba$_{2-x}$Ca$_2$Cu$_3$O$_{6+\delta}$(Tl1223)(Ref. \cite{iyox}) as representative tri-layer system, since no systematic study has ever been done for Bi2223. In Fig. \ref{fig:2}(b), $T_c$ is normalized by the value $T_{c0}$, which is obtained by extrapolating $T_c$ to $\sigma ^2 (r, x)$=0 for each system (see Fig. \ref{fig:5}). The $T_c$-degradation rate against $\sigma ^2 (r, x)$ is found to be roughly 1\ :\ 1/2\ :\ 1/3, for $n$=1, 2, and 3, respectively. The result of $T_c$ degradation against $\sigma ^2$ combined with the result of the residual resistivity in Fig. \ref{fig:2}(a) leads to the ``universal" relationship between $T_c$ degradation and residual resistivity per plane, shown in Fig. \ref{fig:2}(c). Then, we conclude that the charge carriers or the electronic state in the CuO$_2$ plane is affected predominantly by the disorder in the adjacent SrO block and $T_c$ decreases in response to the disorder-induced change in the electronic state.

\subsection{Enhancing $T_c$}

\begin{figure}[t!]
\includegraphics[width=0.9\columnwidth ]{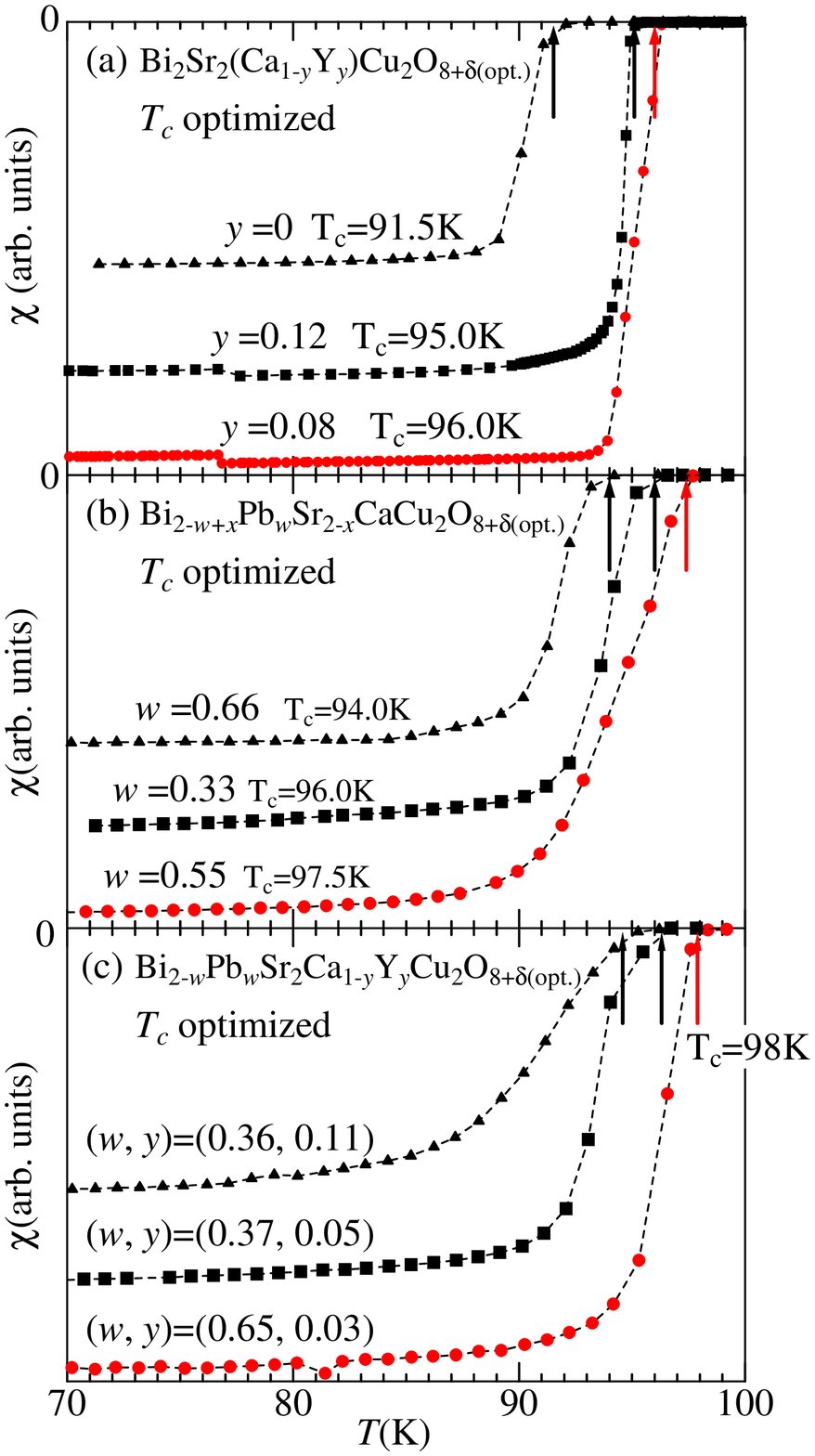}
\caption{\label{fig:3}(a) Superconducting transition temperature measured by magnetic susceptibility for Y-substituted Bi2212. For each Y content $y$ shown are the data optimized by annealing. $T_c$ is maximized by changing excess oxygen content $\delta $ by suitable annealing: for $y$=0, 750$^{\mathrm{o}}$C annealing in atmosphere; for $y$=0.08, 400$^{\mathrm{o}}$C Ar-flow annealing; and for $y$=0.12, without annealing.
(b) Change in $T_c$ for Pb-substituted Bi2212. For each sample, $T_c$ is maximized by adjusting the excess oxygen content $\delta $ under the following annealing conditions: for $w$=0.33, 600$^{\mathrm{o}}$C annealing under N$_2$ flow; for $w$=0.55, 650$^{\mathrm{o}}$C annealing in vacuum; and for $w$=0.66, 550$^{\mathrm{o}}$C annealing under N$_2$ flow. 
(c) Enhancement of $T_c$ with both Y and Pb substitutions in Bi2212. For each sample, ($w$, $y$), $T_c$ is maximized by adjusting excess oxygen content. As the Pb-substituted crystals are naturally overdoped, the annealing was made under reducing atmosphere, and the oxygen content was varied by changing the annealing ($T_A$). The optimal conditions are: for ($w$, $y$)=(0.34, 0.11), 400$^{\mathrm{o}}$C annealing under N$_2$ flow; for ($w$, $y$)=(0.37, 0.05); 650$^{\mathrm{o}}$C annealing in vacuum; and for ($w$, $y$)=(0.65, 0.03), 425$^{\mathrm{o}}$C annealing under N$_2$ flow.}
\end{figure}

\indent Because the Bi/Sr-site mixing easily occurs in Bi2212 and decreases $T_c$, an effective way to improve $T_c$ is to minimize the Bi/Sr-site mixing. The best way to do this would be replacing Ba$^{2+}$ for Sr$^{2+}$, but we cannot make use of this way because Ba in combination with Bi tends to form impurity phases such as BaBiO$_3$. Then, the second best way is to minimize disorder in the SrO block at the expense of disorder in other building blocks, either Bi$_2$O$_2$ or Ca block, or both blocks because we know that the effect of disorder in these blocks on $T_c$ is very small. To do this, we substitute Pb for Bi site in the Bi$_2$O$_2$ block and Y for Ca site in the Ca block. As Pb$^{2+}$ substitution for Bi$^{3+}$ increases the hole density in the CuO$_2$ planes, whereas Y$^{3+}$ substitution for Ca$^{2+}$ decreases it, these substitutions have to be done to the extent that one can compensate for the change in hole density by decreasing/increasing excess oxygen content near Bi$_2$O$_2$ block. In what follows, we show how $T_c$ changes with these processes.

\indent Figure \ref{fig:3}(a) shows how $T_c$ changes by substituting Y for Ca site, Bi$_2$Sr$_2$Ca$_{1-y}$Y$_y$Cu$_2$O$_{8+\delta(opt)}$, with $\delta $ adjusted to give maximum $T_c$ (or to keep the hole density constant). $T_c$ increases up to $y$=0.08, but stops with further substitution of Y. The reason why substituting Y increases $T_c$ is supposed to be that Y$^{3+}$ makes [CuO$_2$-Ca-CuO$_2$] bilayers more electropositive and hence drives Bi$^{3+}$ away from the neighboring SrO blocks. In fact, the ICP analyzed compositions (Table I) of Bi and Sr in the Y-substituted crystal are close to the ideal values, evidencing that the Bi/Sr-site mixing is considerably reduced. However, too much Y substitution ($y>$ 0.08) does not improve $T_c$ any more probably because the decrease in hole density cannot be compensated for by oxygen annealing.

\indent Next, we substitute Pb for Bi in the form Bi$_{2-w+x}$Pb$_w$Sr$_{2-x}$CaCu$_2$O$_{8+\delta }$ [Fig. \ref{fig:3}(b)]. Pb substituted for Bi is considered to take the Pb$^{2+}$ ionic state which donates a hole into the CuO$_2$ planes. Then, $T_c$ is maximized or the hole density is kept at the optimal value by reducing excess oxygen content $\delta $. $T_c$ goes up to $T_c$=97.5K until $w$=0.55. Another known effect of Pb substitution is to relax the structural supermodulation in the BiO planes running along b axes~\cite{sugimoto2},~\cite{motohashi}. However, according to the above analysis, diminishing the supermodulation would have minor effect on $T_c$ because of its location far away from the CuO$_2$ planes. Rather, we ascribe a reason for increased $T_c$ as the decrease in the total amount of Bi, which results in the decrease in Bi content in the SrO block. The ICP analysis gives evidence for this trend (Table I). We also have to take into account possible Pb/Sr-site mixing. However, fortunately the ionic radius of Pb$^{2+}$ is nearly equal to that of Sr$^{2+}$, and the valence of Pb$^{2+}$ is as the same as that of Sr$^{2+}$, so we speculate that Pb/Sr-site mixing is harmless for $T_c$.

\indent From the results shown in Figs. \ref{fig:3}(a) and (b) we know that both Y and Pb substitutions are effective in reducing the Bi/Sr-site mixing. Then, we made simultaneous substitutions of Pb for Bi and Y for Ca site in the form of Bi$_{2-w}$Pb$_w$Sr$_{2}$Ca$_{1-y}$Y$_y$Cu$_2$O$_{8+\delta }$. Figure \ref{fig:3}(c) shows the results of the simultaneous substitution. The maximum $T_c$(=98K, at onset) was achieved for $w$=0.60-0.65, and $y$=0.03-0.05.

\begin{figure}
\includegraphics[width=0.9\columnwidth ]{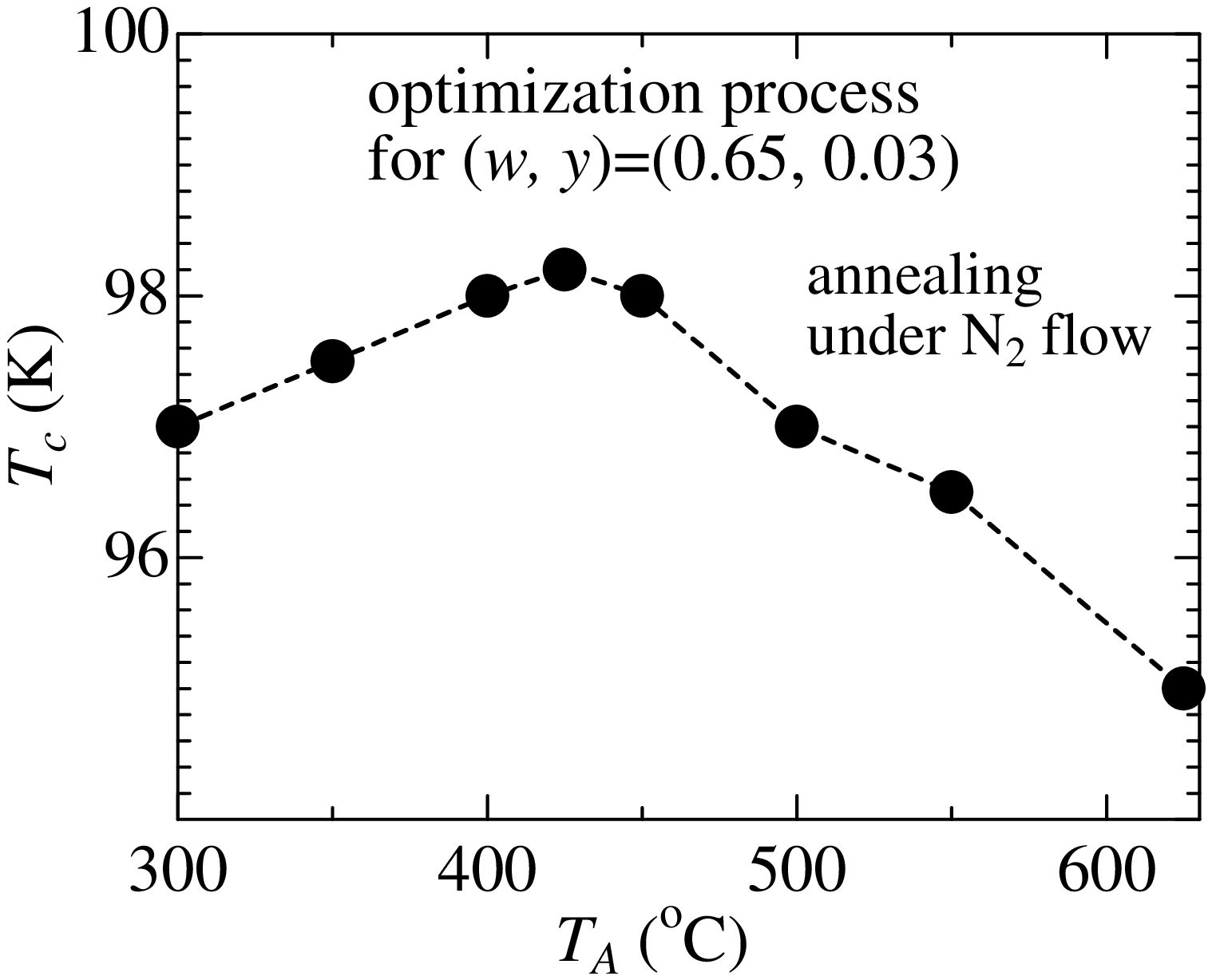}
\caption{\label{fig:4}$T_c$ is plotted against annealing temperature for ($w$, $y$)=(0.65, 0.03). As $T_A$ increases, the excess oxygen content decreases, that is, the doping level changes from over-doping to underdoping, attaining optimal doping at $T_A$=425$^{\mathrm{o}}$C. }
\end{figure}

\section{\label{sec:level3-2}Discussions}

\begin{figure}[b!]
\includegraphics[width=0.9\columnwidth ]{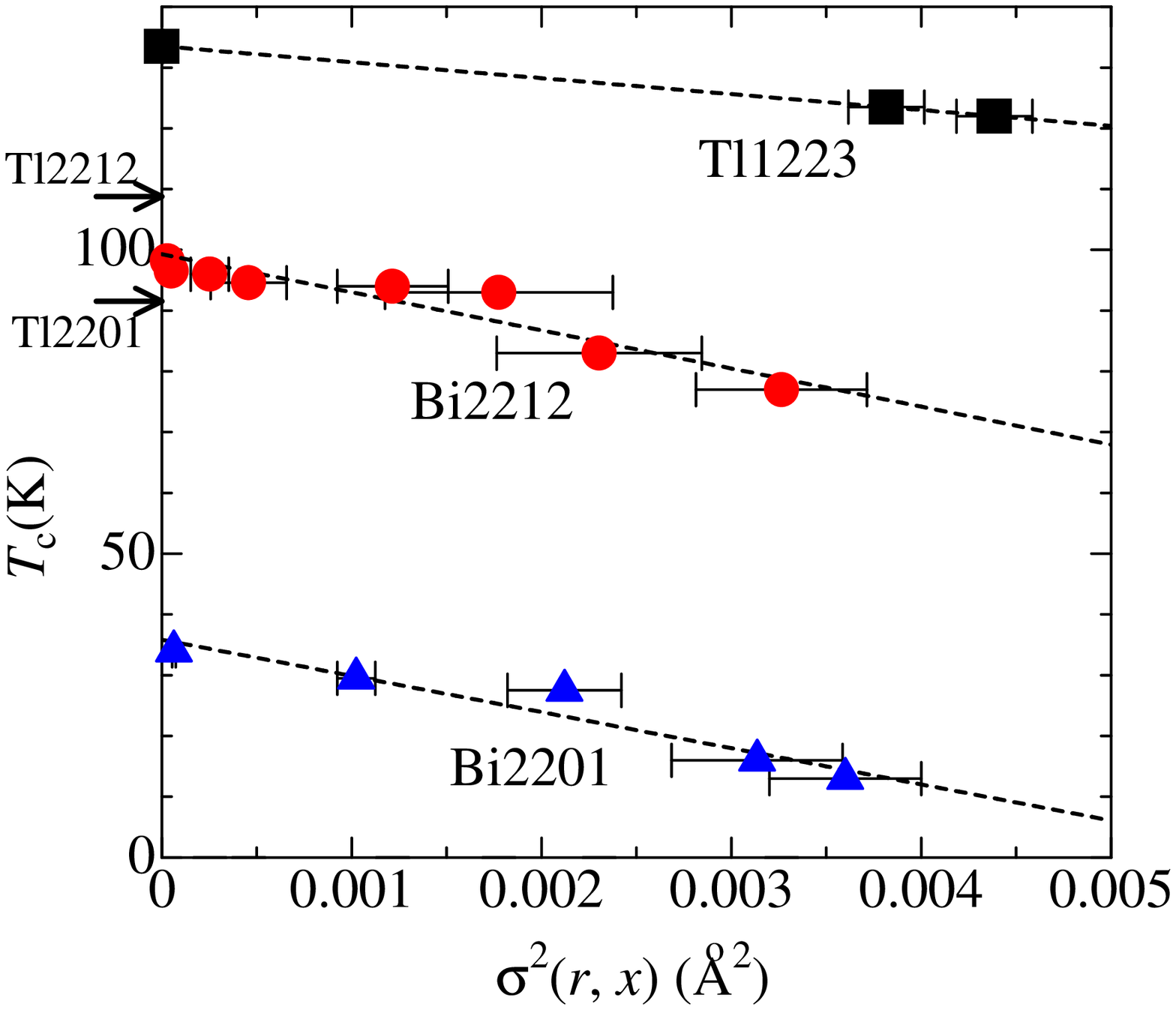}
\caption{\label{fig:5} $T_c$ vs $\sigma ^2(r, x)$ for single-layer, bilayer, and tri-layer cuprates. The arrows indicate the highest $T_c$ values of Tl$_2$Ba$_2$CuO$_{6+\delta}$ (Tl2201) and Tl$_2$Ba$_2$CaCu$_2$O$_{8+\delta}$ (Tl2212), which probably have disorder-free BaO block.}
\end{figure}

\indent Finally, we discuss a maximum $T_c$ value of Bi2212. $T_{c0}$ in Fig. \ref{fig:2}(b) is determined by extrapolating the $T_c$ vs $\sigma ^2 $ plot to $\sigma ^2 $=0 (shown in Fig. \ref{fig:5}). $T_{c0}$ is, in this sense, an upper limit of $T_c$ realized when the disorder in the SrO or BaO block is completely diminished. $T_c$ =98K achieved in the present work should be very close to the limit $T_{c0}$. Although the sample is optimally doped and the disorder in the SrO block is minimized (as also suggested by the ICP analysis listed in Table I), there remains disorder in the Ca and near the Bi$_2$O$_2$ blocks. However, in view of the negligibly small residual resistivity produced by these types of disorders, we expect that the $T_c$ reduction would be very small due to these disorders. In this regard, $T_c$=98.0$\pm$0.5K would be maximum $T_c$ attainable in Bi2212.

\indent In the case of Bi2212, when the Bi/Sr-site mixing is suppressed, the charge disorder is also reduced. This is not the case with Bi2201. $\sigma ^2\sim $ 0 is attained for Ln=La with least mismatch in the ionic radius with that of Sr$^{2+}$. However, because the ionic state of Ln is Ln$^{3+}$ (as compared with Sr$^{2+}$), disorder due to this charge mismatch remains even in the ``optimized'' La-Bi2201. It was reported that $T_c$ increases to $\sim $ 40K when the content of substituted Ln is reduced as much as possible by partial substitution of Pb for Ln/Bi~\cite{arao}. Nevertheless, $T_c$ is much lower than $T_c$=90-98K realized in the isostructural Tl2201 and other single-layer systems. This suggests that the charge disorder at the Sr site equally or more seriously reduces $T_c$.

\indent If $T_c$ of 98K is maximum $T_c$ value of Bi2212, then the question is why it is lower by 10K than the maximum $T_c$ [$\sim $110K (Ref. \cite{shimakawa})] of isostructural Tl2212. We infer that other factors might be responsible for the $T_c$ difference between the two systems. A plausible candidate is the distance of an apical oxygen from the CuO$_2$ plane. The apical-oxygen distances are $d \sim$ 2.43 \AA ~ for Bi2212, $d \sim$ 2.70\AA ~ for Tl2212, and $d \sim$ 2.80 \AA ~ for another bi-layer compound [HgBa$_2$CaCu$_2$O$_{6+\delta}$ (Hg1212)] with highest $T_c$ =127K~\cite{chu}. Apparently, the maximum $T_c$ values are correlated with the apical-oxygen distance, which is also suggested by the empirical relationship based on the local-density approximation (LDA) band calculation~\cite{pavarini}. If we could develop any method to increase the apical-oxygen distance, $T_c$ of Bi2212 would be enhanced to the levels of Tl2212 and of Hg1212.



\indent We speculate that the disorder effect in the SrO block might be intimately connected with the local tilt of the CuO$_6$ octahedron or CuO$_5$ pyramid and/or the local change in the apical-oxygen distance. Local distortions, such as the tilt of CuO$_6$ octahedra, would lead to local modulation of transfer integrals (overlap integrals of the electronic wave functions of neighboring atoms) and/or superexchange interaction between neighboring Cu spins in the CuO$_2$ plane. These modulations would result in the modulation of local pairing interactions or locally stabilize some orders competing with the superconducting order, as suggested by the recent scanning-tunneling-microscopy (STM) observations~\cite{kinoda}-~\cite{slezak} and their theoretical interpretations~\cite{andersen}-~\cite{mori}.

\section{\label{sec:level4}Summary}
We examined the effect of disorder at various out-of-plane lattice sites on $T_c$ in the prototypical high-$T_c$ cuprate Bi$_2$Sr$_2$CaCu$_2$O$_{8+\delta}$. From the $T_c$-degradation rate and the magnitude of induced residual resistivity and also  from the comparison with the results for single-layer and tri-layer Bi-cuprates, we conclude that the disorder in the SrO block substantially reduces $T_c$, while disorder in the Bi$_2$O$_2$ and Ca blocks has very small effect on $T_c$.
The grown crystals of Bi2212 normally contain disorder in the SrO block due to Bi/Sr-site mixing and thus their $T_c$ values are appreciably reduced. So, the effective way to enhance $T_c$ of Bi2212 is reducing Bi/Sr-site mixing as much as possible, and $T_c$=98K is achieved by simultaneous substitutions of Pb for Bi and Y for Ca. The present result gives a good lesson on increasing cuprate $T_c$ based on detailed investigation of the factors that influence $T_c$.


\section*{ACKNOWLEDGMENTS}
We thank Prof. K. Hirota and Dr. T. Suzuki of ISSP, University of Tokyo for their help in the ICP analysis, and for Dr. T. Fujii, cryogenic center, University of Tokyo, for his technical support in resistivity measurement.This work was supported by Grants-in-Aid for scientific Research and the 21st-Century Centers of Excellence Program from the Ministry of Science and Education (Japan) and form the Japan Society for the Promotion of Science.







\end{document}